\documentstyle[12pt]{article}
\begin{document}
{\sf
\title{
{\normalsize
\begin{flushright}
CU-TP-1111\\
RBRC-~~415
\end{flushright}}
A Possible Origin of Dark Energy\thanks{This research was
supported in part by the U.S. Department of Energy Grant
DE-FG02-92ER-40699 and by the RIKEN-BNL Research Center,
Brookhaven National Laboratory}}

\author{
T. D. Lee\\
{\small \it Physics Department, Columbia University, New York, NY 10027}\\
{\small \it China Center of Advanced Science and Technology (CCAST)}\\
{\small \it (World Laboratory), P.O. Box 8730, Beijing 100080, People's Republic of China}\\
{\small \it RIKEN BNL Research Center (RBRC), Brookhaven National Laboratory, Upton, NY 11973}}
\maketitle

\begin{abstract}

We discuss the possibility that the existence of dark energy may
be due to the presence of a spin zero field $\phi(x)$, either
elementary or composite. In the presence of other matter field,
the transformation $\phi(x)\rightarrow \phi(x)~+$ constant can
generate a negative pressure, like the cosmological constant. In
this picture, our universe can be thought as a very large bag,
similar to the much smaller MIT bag model for a single nucleon.

\end{abstract}
\vspace{2cm}

~~~~PACS{:~~98.90.+s,~~12.39.Ba}

\newpage

We assume the existence of a spin zero field $\phi(x)$, like
$\sigma(x)$ in the $\sigma$-model, or the Higgs field in the
Standard Electro-weak Model, or any composite made of other
nonzero spin fields. Let $\phi_{vac}$ be the vacuum expectation
value of $\phi(x)$. Consider an idealized state of a single
particle $i$ with an inertia mass $m_i$. Define the coupling $g_i$
by
\begin{eqnarray}\label{e1}
m_i=g_i~\phi_{vac}.
\end{eqnarray}
The transformation
\begin{eqnarray}\label{e2}
\phi(x) \rightarrow \phi(x)+c,
\end{eqnarray}
where $c$ is a constant, changes both $\phi_{vac}$ and $m_i$:
\begin{eqnarray}\label{e3}
\phi_{vac} \rightarrow \phi_{vac}+c
\end{eqnarray}
and
\begin{eqnarray}\label{e4}
m_i \rightarrow m_i+g_i~c.
\end{eqnarray}
Set $c=-\phi_{vac}$; (\ref{e3}) and (\ref{e4}) become
\begin{eqnarray}\label{e5}
\phi_{vac} \rightarrow 0
\end{eqnarray}
and
\begin{eqnarray}\label{e6}
m_i \rightarrow 0.
\end{eqnarray}
In any non-linear field theory, we can always construct a
composite spin $0$ field $\phi(x)$; therefore, there always exists
a physical state in which the inertia mass of any particles
(excluding $\phi$ itself) is zero. For a single stable particle
state of nonzero $m_i$, by definition the corresponding state of
zero inertia generated by (\ref{e6}) is an excited state. However,
this situation might change for a multi-particle state[1], and
that could be the origin of dark energy (cosmological constant),
as we shall discuss.

A concrete example is the following interpretation of the MIT bag
model[2,3], in which one postulates an energy density function
$U(\phi)$ with
\begin{eqnarray}\label{e7}
U(\phi_{vac})= 0.
\end{eqnarray}
Under the transformation (\ref{e5}) the corresponding change in
$U(\phi)$ is
\begin{eqnarray}\label{e8}
U(\phi_{vac})= 0\rightarrow U(0)\equiv p.
\end{eqnarray}
Thus, the energy $E$ of a bag of radius $r$, containing $N$
quarks, with the expectation value of $\phi$ being zero inside the
bag and $\phi_{vac}$ outside is given by
\begin{eqnarray}\label{e9}
E-\frac{4\pi}{3}r^3p=N\frac{2.0428}{r},
\end{eqnarray}
in which we have to {\it subtract} an amount
\begin{eqnarray}\label{e10}
\frac{4\pi}{3}r^3p
\end{eqnarray}
from $E$ in order to equate the resultant with the matter energy
of quarks. Let $\rho_E$ and $\rho_M$ be the corresponding energy
densities defined by
\begin{eqnarray}\label{e11}
\rho_E \equiv E\Bigm/\frac{4\pi}{3}r^3
\end{eqnarray}
and
\begin{eqnarray}\label{e12}
\rho_M \equiv N\frac{2.0428}{r}\Bigm/\frac{4\pi}{3}r^3.
\end{eqnarray}
Setting
\begin{eqnarray}\label{e13}
\frac{\partial E}{\partial r}=0,
\end{eqnarray}
we find
\begin{eqnarray}\label{e14}
\frac{p}{\rho_M}=\frac{1}{3}~~~~{\sf
and}~~~~~\frac{p}{\rho_E}=\frac{1}{4}.
\end{eqnarray}

We now assume our universe to be a large bag of radius $R$; its
energy ${\cal E}$ can be approximately described by a similar
formula (\ref{e9}), with $p$ replaced by the energy density
$\rho_{\Lambda}$ due to the cosmological constant:
\begin{eqnarray}\label{e15}
{\cal E}-\frac{4\pi}{3}R^3\rho_{\Lambda}\cong {\cal M}/R,
\end{eqnarray}
where ${\cal M}$ is a constant and ${\cal M}/R$  representing the
matter energy, assumed to be dominated by its long wave length
component. Setting $\partial {\cal E}/\partial R=0$, we find the
total energy density
\begin{eqnarray}\label{e16}
\rho_{{\cal E}}={\cal E}\Bigm/\frac{4\pi}{3}R^3
\end{eqnarray}
is related to $\rho_{\Lambda}$ by
\begin{eqnarray}\label{e17}
\rho_{\Lambda}/\rho_{{\cal E}} \cong \frac{1}{4}.
\end{eqnarray}
Likewise, defining the matter-energy density of the universe by
\begin{eqnarray}\label{e18}
\rho_{{\cal M}} \equiv \frac{{\cal M}}{R}\Bigm/\frac{4\pi}{3}R^3,
\end{eqnarray}
we derive
\begin{eqnarray}\label{e19}
\rho_{\Lambda}/\rho_{{\cal M}} \cong \frac{1}{3}.
\end{eqnarray}
Taking the present value of $\rho_{\Lambda}$[4] to be
\begin{eqnarray}\label{e20}
\rho_{\Lambda} \cong 3 \times10^{-6} GeV/cm^3,
\end{eqnarray}
we find
\begin{eqnarray}\label{e21}
\rho_{{\cal M}} \cong 9 \times10^{-6} GeV/cm^3
\end{eqnarray}
and
\begin{eqnarray}\label{e22}
\rho_{{\cal E}} \cong 1.2 \times10^{-5} GeV/cm^3;
\end{eqnarray}
both are of the same order of magnitude as the critical density
$\rho_c$ of the universe. Considering our over-simplification of
neglecting the non-Euclidean geometrical effect, the shorter
wavelength contribution of the matter energy and the dynamical
effect of our universe expansion, the above order of magnitude
agreement is very encouraging to the suggestion that the negative
pressure postulated by Poincar\'{e}[5], Dirac[6], MIT-bag[2] and
the cosmological constant[7] can all be attributed to the
existence of an effective scalar field which is kinematically
connected to the inertia of all matter.

The pressure of the MIT bag is exerted by the physical space
outside the bag; likewise, the negative pressure described by the
cosmological constant may also be due to the physical space
outside the horizon of our universe. Through relativistic heavy
ion collisions (RHIC), we can change the bag-pressure $p$ and the
quark energy density $\rho_M$; likewise, by examining carefully
the dynamical change of matter energy within our universe, we may
also gain insight to the universe that lies outside our horizon.
In this picture, most likely our universe is not a self-contained
system, and the "cosmological constant" may actually be a
dynamical variable, related to $\phi(x)$. An analysis of these
interesting possibilities will be given in a separate publication.

The author wishes to thank Miklos Gyulassy for discussions and
Charles Baltay for a private communication.

\vspace{1cm}

 \noindent {\large \bf References}

\noindent
[1] T. D. Lee and G. C. Wick, Phys. Rev. D9(1974)2291\\
~[2] A. Chodos, et al., Phys. Rev. D9(1974)3471\\
~[3] T. D. Lee, Particle Physics and Introduction to Field Theory,
p569

\noindent~~~~~~~~~~ Harwood Academic Publishers, 1981\\
~[4] C. Baltay, private communication \\
~[5] A. Einstein, The Meaning of Relativity, p106

\noindent~~~~~~~~~~ Princeton University Press, 1950\\
~[6] P. A. M. Dirac, Proc. R. Soc. A268(1962)57  \\
~[7] A. Einstein, The Meaning of Relativity, p111

\noindent~~~~~~~~~~ Princeton University Press, 1950\\

\end{document}